\documentstyle[preprint,aps]{revtex}

\begin{document}
\title{Tolman's energy of a stringy charged black hole}
\author{S.~S.~Xulu\footnotemark[1] 
\footnotetext[1]{Dept of Applied Mathematics, University of Zululand,
Private Bag X1001, 3887 KWA-DLANGEZWA, South Africa. 
E-mail: ssxulu@pan.uzulu.ac.za}}
\maketitle
 \begin{abstract}
 Virbhadra and Parikh studied the energy distribution associated with 
 stringy charged black hole in Einstein's prescription. We study the
 same using Tolman's energy-momentum complex and get the same result as
 obtained by Virbhadra and Parikh. The entire energy is confined inside
the black hole.
 \end{abstract}
The energy-momentum localization has been one of the most important
and challenging research topics in general theory of relativity. This
subject has fascinated many renowned scientists (see M\o ller 1958,
Hawking 1968, Penrose 1982, and Rosen 1994 and references therein).
Following the Einstein energy-momentum complex many definitions
of energy,  momentum, and angular momentum for a general relativistic
system  have been proposed (see in Brown and York 1993, Aguirregabiria et al.
1996).  The physical interpretation of these
nontensorial energy-momentum complexes have been questioned by
a number of physicists, including Weyl, Pauli and Eddington
(see in Chandrasekhar and Ferrari 1991).  There prevails suspicion
that different energy-momentum complexes could give different
energy distributions in a given spacetime. To this end Virbhadra and
his collaborators have  considered many spacetimes and have shown that
several energy-momentum complexes give the same and acceptable result
for a given spacetime. Virbhadra (1990) considered the Kerr-Newman metric
and carried out calculations up to the third order of the rotation parameter.
Cooperstock and Richardson (1991) extended his investigations up to the
seventh order of the rotation parameter and reported that several definitions
yield the same result.  Virbhadra (1992) carried out calcualtions
for the Vaidya (radiating Schwarzschild) metric and found that several
definitions give the same result.  Virbhadra and Parikh (1994), after
obtaining a conformal scalar dyon black hole solution, calculated the
energy distribution with this black hole and found reasonable result.
Chamorro and Virbhadra (1995) studied
the Bonnor-Vaidya spacetime and found that energy-momentum complexes 
give the same result as obtained by Tod using the Penrose definition
(see for details in Aguirregabiria et al. 1996). Rosen and Virbhadra
(1993) and Virbhadra (1995) investigated the well-known Einstein-Rosen
spacetime and  found that several energy-momentum complexes give the
same result for the energy and energy current densities.
 Aguirregabiria et al. (1996) showed that several well-known energy-momentum
complexes coincide  for any Kerr-Schild class metric.  These are definitely 
encouraging results.

For the Reissner-Nordstr\"{o}m metric, several definitions of energy
give
\begin{equation}
E(r) = M - \frac{Q^2}{2r}
\label{eqn1}
\end{equation}
(see in Tod 1983, Hayward 1994, Aguirregabiria et al. 1996).
Thus the energy is both in its interior and exterior. 
There is considerable difference in the solutions in the Einstein-Maxwell
theory and the low-energy string theory. Virbhadra and Parikh (1993)
deduced, using Einstein's prescription, that gravitational energy of a stringy 
charged black hole is given by $E = M$.
and thus the energy is confined to the interior of the black hole.  
It is worth investigating whether or not other definitions of energy
give the same result as obtained by them.
Throughout this paper we use $G=1$ and $c=1$ units and follow the
convention that Latin indices take values from $0$ to $3$ and Greek
indices take values from $1$ to $3$.

For a static, spherically symmetric charged black hole in low-energy string
theory, the line element is given by (Garfinkle et al. 1991)
\begin{equation}
ds^{2} =(1- \frac{2M}{r}) dt^2-(1- \frac{2M}{r})^{-1} dr^2
-(1- \frac{\alpha}{r}) r^2 (d \theta^2 + sin^2 \theta d \phi^2) ,
\label{eqn2}
 \end{equation}
where
 \begin{equation}
~~ \alpha = Q^2 \frac{exp{(-2 \phi_0)}}{M}.
\label{eqn3}
 \end{equation}
$M$ and $Q$ are, respectively, mass and charge parameters; $ \phi_0 $
is the asymptotic value of the dilaton field. 
It is well-known that the energy-momentum complexes give meaningful result
if calculations are performed in quasi-Cartesian coordinates.
The line element $(\ref{eqn2})$  may be transformed to quasi-Cartesian 
coordinates:
\begin{equation}
ds^{2}  = (1-\frac{2M}{r})dt^2 - (1-\frac{\alpha}{r}) (dx^2 + dy^2 + dz^2)
-\frac{(1-2M/r)^{-1}-(1-\alpha /r)}{r^2} (xdx + ydy + zdz)^2 ,
 \label{eqn4}
 \end{equation}
according to                               
 \begin{equation}
r = \sqrt{x^2 + y^2 + z^2}, ~~ \theta =\cos^{-1}\left(\frac{z}{\sqrt{x^2 + y^2 + z^2}}\right),    
~~ \phi = \tan^{-1} (y/x) .
\label{eqn5}
 \end{equation}

The  energy-momentum complex of Tolman (1934) is
\begin{equation}
{\cal{T}}_k{}^i = \frac{1}{8 \pi}U^{ij}_{k,j},
\label{eqn6}
\end{equation}
where
\begin{equation}
U^{ij}_{k} = \sqrt{-g} \left[-g^{pi}( -\Gamma^j_{kp}
               +\frac{1}{2} g^j_{k} \Gamma^a_{ap}
               +\frac{1}{2} g^j_{p} \Gamma^a_{ak})  
           + \frac{1}{2} g^i_{k} g^{pm}( -\Gamma^j_{pm}
               +\frac{1}{2} g^j_{p} \Gamma^a_{am}
               +\frac{1}{2} g^j_{m} \Gamma^a_{ap}\right] .
\label{eqn7}
\end{equation}
${\cal{T}}_0^0$ is the energy density, ${\cal{T}}_0^{\alpha}$ are 
the components of energy current density, and  ${\cal{T}}_{\alpha}^{0}$ are 
the momentum density components.  The energy $E$ is given by the 
expression:
\begin{equation}
E = \int\int\int {\cal{T}}_0^0 dx dy dz
\label{eqn8}\\
\end{equation} 
Using the Gauss theorem (noting that the spacetime under consideration is
static) one has
\begin{equation}
E = \int \int U^{0 \beta}_0  n_{\beta} dS
\label{eqn9}
\end{equation}
$n_{\beta}$ stands for the 3-components of unit vector over an infinitesimal
surface element $dS$.
For the metric given by Eq. $(\ref{eqn4})$ Virbhadra and Parikh (1995) computed the 
determinant of the metric tensor and the  contravariant components of the 
 tensor. To compute energy using Eq. $(\ref{eqn9})$, we 
require the following list of  nonvanishing components  of 
Christoffel symbol:
\begin{eqnarray}
\Gamma^0_{01} &=& \frac{M x}{r^2 (r-2M)},
                       ~~~~~~~~~~~~\Gamma^0_{02} = \frac{M y}{r^2 (r-2M)},
                        \nonumber\\
\Gamma^0_{03} &=& \frac{M z}{r^2 (r-2M)},
                        ~~~~~~~~~~~~\Gamma^1_{00} = \frac{M(r-2M)x}{r^4},
                        \nonumber\\
\Gamma^2_{00} &=& \frac{M(r-2M)y}{r^4},
                        ~~~~~~~~~~\Gamma^3_{00} = \frac{M(r-2M)z}{r^4},                        \nonumber\\
\Gamma^1_{11} &=& x ( a_1 + a_2 x^2 ),
                        ~~~~~~~~~~~~\Gamma^2_{11} = y ( a_3 + a_2 x^2 ),
                        \nonumber\\
\Gamma^3_{11} &=& z ( a_3 + a_2 x^2 ),
                        ~~~~~~~~~~~~\Gamma^1_{22} = x ( a_3 + a_2 y^2 ),
                         \nonumber\\
\Gamma^2_{22} &=& y ( a_1 + a_2 y^2 ),
                        ~~~~~~~~~~~~\Gamma^3_{22} = z ( a_3 + a_2 y^2 ),
                        \nonumber\\
\Gamma^1_{33} &=& x ( a_3 + a_2 z^2 ),
                       ~~~~~~~~~~~~\Gamma^2_{33} = y ( a_3 + a_2 z^2 ),
                       \nonumber\\
\Gamma^3_{33} &=& z ( a_1 + a_2 z^2 ),
                        ~~~~~~~~~~~~\Gamma^1_{12} = y (a_4 + a_2 x^2 ),
                        \nonumber\\
\Gamma^1_{13} &=& z (a_4 + a_2 x^2 ),
                        ~~~~~~~~~~~~\Gamma^2_{21} = x (a_4 + a_2 y^2 ),
                        \nonumber\\
\Gamma^2_{23} &=& z (a_4 + a_2 y^2 ),
                        ~~~~~~~~~~~~\Gamma^3_{31} = x (a_4 + a_2 z^2 ),
                        \nonumber\\
\Gamma^3_{32} &=& y (a_4 + a_2 z^2 ),
                        \nonumber\\
 \Gamma^1_{23} ~~&=& ~~\Gamma^2_{13} ~~= ~~\Gamma^3_{12} ~~=  ~~a_1 x y z ,
\label{eqn10}
\end{eqnarray}
where\\
\begin{eqnarray}
a_1 &=& \frac{1}{2r^4} ( \alpha r + 4 M r - 2 \alpha M 
                        + \frac{2 \alpha r^2}{r-\alpha} ) ,
                        \nonumber\\
a_2 &=& \frac{1}{2 r^6} ( 2\alpha M - 3\alpha r - 4Mr 
                        - \frac{2Mr}{r-2M} - \frac{2\alpha^2 r}{r-\alpha}) ,
                        \nonumber\\
a_3 &=& \frac{1}{2r^4} ( \alpha r + 4 M r - 2 \alpha M ) ,
                        \nonumber\\
a_4 &=& \frac{ \alpha }{2r^2 (r-\alpha)} .
\label{eqn11}
\end{eqnarray}
Further, after straightforward but very lengthy calculations we get 
\begin{eqnarray}
U^{01}_{0} = \frac{2Mx}{r^3}, \nonumber\\
U^{02}_{0} =  \frac{2My}{r^3}, \nonumber\\
U^{03}_{0} =  \frac{2Mz}{r^3}.   
\label{eqn12}
\end{eqnarray}
Now using  $(\ref{eqn12}), n_{\beta}= (x/r, y/r,z/r),$ and $dS = r^2 \sin\theta 
d\theta d\phi$  in $(\ref{eqn9})$, we get
\begin{equation}
E = M .
\end{equation}
Thus the entire energy of a charged black hole in low-energy string theory
lies inside the black hole. The ``effective gravitational mass'',
given by Eq. $(13)$, that a netral
test particle  experiences  is always positive. This is quite different
from the case of the Reissner-Nordstr\"{o}m metric (for $r < Q^2/2M$ the 
``effective gravitational mass'', $E(r)$, is negative). Further, it is  worth noting 
that we, using the Tolman energy-momentum complex, got the same result
as obtained by Virbhadra and Parikh (they used the Einstein energy-momentum
complex). Our result supports  importance of  energy-momentum
complexes.

\acknowledgments

Thanks are due to  K. S. Virbhadra for his guidance.

\end{document}